\documentstyle[prl,tighten,aps,epsf,graphics,multicol,amstex]{revtex}
\begin{document}

\draft

\title{The Holevo bound and Landauer's principle}
\author{Martin B. Plenio}
\address{Optics Section, Blackett Laboratory, Imperial
College, London SW7 2BZ, United Kingdom;\\
m.plenioaic.ac.uk, fax: +44 (0)171 823 8376}

\date{\today}
\maketitle

\begin{abstract}
Landauer's principle states that the erasure of information 
generates a corresponding amount of entropy in the environment. 
We show that Landauer's principle provides an intuitive basis
for Holevo bound on the classical capacity of a quantum channel.
\end{abstract}


\begin{multicols}{2}
\narrowtext

\section{Introduction}
Landauer's principle \cite{Landauer} states that the erasure of
classical information leads to an increase in the entropy of 
the environment by at least the same amount. The importance of
this principle is the insight that it is not the act of obtaining
information that necessarily generates heat and therefore entropy, but 
the erasure of information. This profound insight has led to the 
resolution of the problem of Maxwell's demon by Bennett \cite{Bennett}. 
Here the missing entropy is generated when the demon's memory is
erased. Clearly Landauer's principle provides a bridge that links classical 
information theory to thermodynamics. Recently, however, it has 
been shown that one can also use Landauer's principle to connect
the entropy of erasure and the efficiency of entanglement purification 
\cite{Vedral}.

Here we demonstrate that there is also a connection between Landauer's 
principle and the Holevo bound which limits the classical capacity 
of a quantum channel. In classical communication via quantum states,
we encode classical signals in possibly mixed, non-orthogonal quantum 
states and send those to a receiver. The receiver then has the task 
to deduce from those quantum states the original classical message.
The maximal information that the receiver can obtain is limited by the 
Holevo bound \cite{Holevo1}. It has also been shown that this bound can 
be achieved asymptotically \cite{Holevo2}.

The purpose of this paper is to understand the Holevo bound in terms 
of Landauer's principle. This approach is more intuitive than the 
unavoidably very technical, rigorous mathematical proofs. The
hope is that this new approach may stimulate ideas that may lead to a 
better understanding of the more intricate problem of the quantum capacity 
of a quantum channel, a problem which is not yet fully understood
(see however \cite{Schumacher}). 

This paper is organized as follows. In the next section we will
present a general method for the erasure of information which was
introduced by Lubkin \cite{Lubkin} and later generalized by Vedral 
\cite{Vedral} to non-commuting variables. In section III we will 
first briefly introduce the Holevo bound and then proceed to show 
how the Holevo bound can be obtained from Landauer's principle. 

\section{How to erasure information?}
\label{Lubkinsec}
Landauer's principle states that the erasure of a given 
amount of information generates at least the same amount 
of entropy. In the following we will present a physical 
scheme for the erasure of information (due to Lubkin 
\cite{Lubkin}) in its form for non-commuting states as described
in \cite{Vedral}. This scheme has the nice feature that it 
can easily be made optimal in the sense that the entropy 
of erasure can be made equal to the amount of information 
that has been erased.

Let us consider a measurement apparatus ${\cal M}$ that is
initially in a pure state (e.g. the ground state of ${\cal M}$)
and then interacts with a system. This interaction results in a 
measurement, i.e. after the interaction the apparatus will be in 
one of a set of pure states $\{|m_i\rangle\}$ with probabilities 
$\{p_i\}$. The amount of classical information the apparatus 
${\cal M}$ has acquired during the measurement is given by 
the mutual information between system and apparatus which turns out be
equal to the von Neumann entropy $S(\rho)$ where $\rho=\sum_{i} p_i  |m_i\rangle\langle m_i|$ is the average state of the apparatus
\cite{remark}. If we want to reuse the 
apparatus ${\cal M}$ for another measurement, then this information 
has to be erased, i.e. we have to return ${\cal M}$ to its original 
pure state. This can be done with arbitrary precision by placing 
${\cal M}$ into contact with a heat bath of temperature $T$ such 
that in thermal equilibrium the apparatus ${\cal M}$ will be 
essentially in its ground state (This can always be achieved
by using an apparatus ${\cal M}$ with a sufficiently large level
spacing). Note, that we do not discard the apparatus and replace
it by some other system. In particular we do not dump it into the 
heat bath, as the number of particles in the heat bath has to be 
preserved. Only energy is exchanged between the apparatus and the 
heat bath. 

For an arbitrary state $\omega$ we can always chose the temperature
$T$ of the heat bath such that in thermal equilibrium the state 
of the apparatus is described by the Boltzmann distribution
\begin{equation}
	\omega = Z^{-1} e^{-\beta H} \;\; ,
\end{equation}
where $\beta=1/kT$, $H$ is the Hamilton operator of the apparatus
and $Z=tr e^{-\beta H}$. To erase the apparatus we place it into 
contact with the heat bath. The total change of entropy of erasure
is given by the sum of the changes of entropy both in the measurement 
apparatus as well as the heat bath. After the measurement, and prior 
to the erasure, the apparatus is 
in one of the pure states $|m_i\rangle$. Therefore it will always 
increase its entropy by evolving into the state $\omega$ given in
Eq. (1). This change of entropy of the apparatus is given by
\begin{equation}
	\Delta S_M = S(\omega) \;\; .
	\label{change1}
\end{equation}
Now we need to evaluate the change of entropy of the heat bath 
$\Delta S_{B}$. The easiest way to do so is by determining the 
heat change of the bath. This is just the negative of the 
change of heat in the measurement apparatus ${\cal M}$, for which we find
\begin{eqnarray}
	kT \Delta S_B &=&  tr\{H (\rho - \omega) \}  \nonumber \\
	&=& -kT\, tr \{ (\rho - \omega) \log Z\omega \} \nonumber \\
	&=& -kT\, tr \{ (\rho - \omega) \log \omega \} \label{change2} \;\; .
\end{eqnarray}
Therefore, the total change of entropy $\Delta S_{tot}$ of measurement 
apparatus and heat bath together is given by
\begin{eqnarray}
	\Delta S_{tot} &=& \Delta S_M + \Delta S_B  \nonumber \\
	&=& S(\omega) - tr \{ (\rho - \omega) \log \omega \} \nonumber \\
	&=& - tr \{ \rho \log \omega \} \label{change3} \;\; .
\end{eqnarray}
We can see that the amount of erased information $S(\rho)$ is never 
larger than the entropy of erasure, i.e.  
\begin{equation}
	S(\rho) \le \Delta S_{tot} = - tr \{ \rho \log \omega \}
	\label{change4}
\end{equation}
because the relative entropy 
$S(\rho||\omega) = tr \{ \rho \log\rho - \rho\log\omega \}$ is always 
positive. If the temperature of the heat bath is chosen such that for the 
thermal equilibrium state of the apparatus in Eq. (1) we have $\omega=\rho$, 
then the entropy of erasure is exactly equal to the information that has been
erased from the apparatus and the erasure is optimal. 

\section{The Holevo bound}
\subsection{Classical Information via a quantum channel}
\label{Classical}
The transmission of classical information via a quantum channel 
proceeds in the following way. Initially the sender, Alice, holds 
a long classical message. She encodes letter $i$ (which appears 
with probability $p_i$) of this message into a possibly mixed quantum 
state $\rho_i$. These quantum states are handed over to the receiver, 
Bob, who then has the task to infer Alice's classical message from 
these quantum states. The upper bound for the capacity for such a 
transmission, i.e. the information $I$ that Bob can obtain about Alice's 
message per sent quantum state, is given by the Holevo bound \cite{Holevo1}
\begin{equation}
	I \le I_H = S(\rho) - \sum_i p_i S(\rho_i) \;\; .
	\label{Holevo}
\end{equation}
In fact, equality can be achieved for large message-blocksizes as has 
been proven by Holevo \cite{Holevo2}. The aim of the next section is 
to show how one can justify Holevo's bound from the assumption of 
the validity of Landauer's principle.

\subsection{Holevo's bound from Landauer's principle}
The idea behind the derivation of the Holevo bound from Landauer's
principle is to determine an upper bound on the entropy that is 
generated when Bob erases the information that the message system carries 
in its state $\rho_i$. 
In this way we directly obtain an upper bound on the information 
received by Bob. To this end we consider different ways for erasing 
the information that Alice has originally encoded. 
The two methods of erasure are schematically presented in Fig. 1. 
Step 2 of procedure (2) corresponds to the erasure of Bob's information. 
In the following the accompanying entropy of erasure will be computed 
as the difference of the entropies of erasure in procedure (1) and 
the first step of procedure (2).
\begin{figure}[hbt]
\begin{center}
\epsfxsize8.0cm
\centerline{\epsfbox{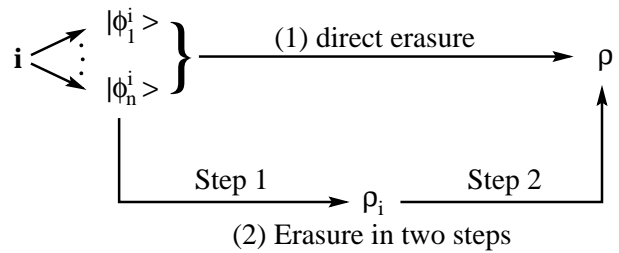} }
\end{center}
\vspace*{-0.75cm}
\caption{\label{Fig1} A letter of a classical message is encoded with probability
$r^i_{\alpha}$ into a pure quantum state $|\phi_{\alpha}^{i}\rangle$. The information
about the original message contained in this encoding can be deleted in two
ways. (1) directly by placing the systems into contact with a heat bath in state
$\rho$, or (2) by first using heats baths in state $\rho_i$ and then a heat bath in
state $\rho$.}
\end{figure}
Let us begin by assuming that Alice initially encodes her message
in the following way. Given she wants to send letter $i$, she then 
encodes it into one of the pure states 
$\{ |\phi_{\alpha}^{i}\rangle|\alpha=1,\ldots,N\}$ with probability $r_{\alpha}^i$
such that 
$\rho_i= \sum_{\alpha} r_{\alpha}^i |\phi_{\alpha}^{i}\rangle\langle\phi_{\alpha}^{i}|$.
As she has encoded her message in pure states, the mutual information
between message and quantum states is $S(\rho)\equiv S(\sum_i p_i \rho_i)$.
This is the information that the encoded quantum states contain about 
the original classical message. Now we will erase this information in 
two different but equivalent ways.

\subsubsection{Direct erasure}
A message letter $i$ which appears with probability $p_i$, is encoded
by Alice with probability $r_{\alpha}^i$ in state $|\phi_{\alpha}^i\rangle$.
We will now delete the information encoded in these pure state by 
bringing them into contact with a heat bath. We chose the temperature
of this heat bath such that the thermal equilibrium state of the message system
is $\rho$. This ensures that the erasure is optimal, in the sense that it 
produces the smallest possible amount of heat. 
This procedure is illustrated as part (1) in Fig. 1. Following the 
analysis of Lubkin's erasure in section \ref{Lubkinsec}, the entropy 
of erasure is given by
\begin{equation}
	\Delta S_{er}^{(2)} = -\sum_i p_i tr\{\rho_i\log\rho\} = S(\rho) \;\; .
\end{equation}
Note that all information has been deleted because now every quantum
system is in the same state $\rho$ so that there is no correlation
between the original letter $i$ and the encoded quantum state left 
after the erasure! 
%In the following we will consider a two step procedure for erasure
%that leads to the same final result, i.e. total erasure again in an optimal way.

\subsubsection{Two step erasure}
{\em First step:} We begin by performing a partial erasure on the 
encoded quantum systems (see step 1 of procedure (2) in Fig 1). For 
a fixed $i$ which appears with probability $p_i$, we place the encoded 
pure states into contact with a heat bath. The temperature $T$ of the 
heat bath is chosen such that the thermal equilibrium state of the
message system is $\rho_i$. Again this choice ensures that the erasure is optimal. 
According to our analysis of the Lubkin erasure in section II, the 
entropy of erasure is then found to be
\begin{eqnarray}
	\Delta S_{er}^{(1)} &=& -\sum_i p_i \sum_{\alpha} 
	tr\{ r_{\alpha}^i |\phi_{\alpha}^i\rangle\langle\phi_{\alpha}^i| \log\rho_i\} \nonumber\\
	&=& -\sum_i p_i tr\{\rho_i\log\rho_i\} \nonumber\\
      &=& \sum_i p_i S(\rho_i) \;\; .
\end{eqnarray}
After this first step in the erasure procedure there is still some
information left in the physical systems. The letter $i$ of the 
classical message is correlated with the state $\rho_i$ of the 
quantum system. In fact, this is exactly the situation in which 
Bob is after he received a message which is encoded as in described 
in subsection \ref{Classical}. To obtain a bound on the information 
that Bob is now holding, we need to find a bound on the entropy of 
erasure of his quantum systems.

{\em Second step:} In order to carry out step 2 of procedure (2) we
place each of Bob's systems, which is in one of the states $\rho_i$ 
with probability $p_i$, into contact with a heat bath such that the 
thermal equilibrium state of the message system is $\rho$.
As the average state of the systems is $\rho=\sum_i p_i \rho_i$, we 
expect the erasure to be optimal again. We can see easily (Fig. 1) 
that this second step of erasure, just generates an amount of entropy that 
is the difference between the entropy of erasure of the first procedure and 
that of the first step of the second procedure. Therefore the entropy of 
erasure of Bob's systems which are in one of the states $\rho_i$'s is 
\begin{eqnarray}
	\Delta S_{er}(Bob) &=& \Delta S_{er}^{(2)} - \Delta S_{er}^{(1)} \nonumber \\
	&=& S(\rho) - \sum_i p_i S(\rho_i) \;\; .
\end{eqnarray}
As the largest possible amount of information available to the receiver 
Bob is bounded by his entropy of erasure we have
\begin{equation}
	I \le \Delta S_{er}(Bob) = S(\rho) - \sum_i p_i S(\rho_i) = I_H \;\; .
\end{equation}
Therefore we have obtained the Holevo bound on the information in the 
states $\rho_i$ which appear with probabilities $p_i$.  
While this derivation only establishes the Holevo bound as an upper
bound, one may argue that if Bob's quantum states contain less than 
$I_H$, then we would expect to be able to find an even lower 
entropy of erasure for his message. Of course such an argument for the 
achievability of the Holevo bound cannot replace a full analytical 
proof, but merely forms the basis for a conjecture which is likely to 
be true.

\section{Conclusions}
In this paper we have shown a different way of understanding the 
origin of the Holevo bound on the classical information capacity 
of an encoding using mixed quantum states. While the resulting bound
is of course not new in itself we hope that this approach to
the Holevo bound may help to stimulate new insights into the more 
difficult and yet to be solved question of the quantum capacity 
of a quantum channel.\\

Acknowledgements: The author thanks Vlatko Vedral for discussions on 
the subject of this paper. This work is supported by the EPSRC, 
The Leverhulme Trust, and the European Union TMR Network ERBFMRXCT960066.

\end{multicols}

\begin{thebibliography}{99}
\bibitem{remark} This information can be achieved which follows from 
Schumachers noiseless coding theorem.
\end{thebibliography}
\end{document}